\documentclass[12pt,nofootinbib]{revtex4}
\usepackage{color}
\usepackage{amssymb,amsmath}
\usepackage{graphicx}
\usepackage{epstopdf}

\newcounter{aaa}\newcounter{bbb}
\newenvironment{teor*}[2][{}]{\begin{trivlist}\refstepcounter{aaa}%
\labelsep=0pt\item[\bfseries #2. ]#1}
{\end{trivlist}}

\newcommand{\rmd}{\mathrm{d}}
\newcommand*{\bi}[1]{\boldsymbol{#1}}
\newcommand{\obozn}{\equiv}
\newcommand{\evalat}[3]{\left.#1\right|_{#2}^{#3}}%

\newcommand{\ssy}[5]{#1,    #2 {\bf #3}, #5 (#4)\rlap{.}}

\begin{document}
\title{Is the  Kerr black hole a super accelerator?}
\author{S. Krasnikov}\email{krasnikov.xxi@gmail.com}
\affiliation{Central Astronomical Observatory at Pulkovo, St.Petersburg, 196140, Russia}
\author{M. V. Skvortsova}\email{milenas577@mail.ru}
\affiliation{Peoples� Friendship University of Russia (RUDN University), 6 Miklukho-Maklaya St, Moscow, 117198, Russian Federation}
\begin{abstract}
A number of long-standing puzzles, such as the origin of extreme-energy cosmic rays, could perhaps be solved if  we found a mechanism for effectively transferring energy from  black holes to  particles and, correspondingly, accelerating the latter to (unboundedly, as long as we neglect the back reaction) large velocities. As of today  the only  such candidate mechanism in the case of  the nonextreme  Kerr black hole is colliding a particle that freely falls from infinity  with a particle whose trajectory is subject to some special requirements to fulfil which it has to be  suitably   corrected by auxiliary collisions. In the present  paper we prove that---at least when the relevant particles move in the equatorial plane and experience a single correcting  collision---this mechanism does not work too. The energy of the final collision becomes unboundedly high only when the energies of the incoming particles do.
\end{abstract}
\maketitle
\section{Introduction and conclusion}
Black holes (BHs) are often thought of as one-way membranes and one does not expect them to be sources of high-energy phenomena such as galactic jets or cosmic rays. Penrose   \cite{Penr1,Penr2} was first to find a mechanism
enabling one to extract the rotational energy from a Kerr black hole.  The idea is based on the fact that near the horizons of these BHs particles may have ``negative energies.'' So, one can  split a particle falling in  a   Kerr BH into two pieces one of which  acquires  negative energy, while the other escapes to infinity with energy  greater then that of the initial particle.
Another way of extracting the  energy of a rotating BH is \emph{superradiance}---wave amplification due to scattering from the black hole---discovered  by Zel'dovich and Starobinsky \cite{Zel,Star}.

	In response to Wheeler's idea \cite{Wheeler} that it is the ``energy mining from BHs'' that underlies the galactic jets, Bardeen, Press and Teukolsky \cite{BPT} thoroughly analyzed   the Penrose process for stars tidally disrupting near a Kerr black hole. They came to the conclusion that   for  real astrophysical objects no  significant energy gain can be obtained in this process. The same turned out  to be true in the general case: as was shown by Wald \cite{Wald} the energy of the outgoing particle in the  above-described reaction is roughly the same as it would be in the Minkowski space.

	Shortly thereafter  Piran, Shaham and Katz \cite{Pir} considered the Penrose process in which a  particle instead  of decaying scatters from  another particle. They argued that this new process is much more efficient and
  that the energy gain grows unboundedly as  the BH approaches  the extreme (i.~e., the maximally spinning) Kerr solution.  	 The limiting case  was considered  in 2009 by Banados, Silk, and West (BSW), who found  a pole in the expression for the energy of a two-particle collision in a vicinity of an extreme Kerr BH \cite{bsw}. Their discovery meant that such BHs work as  super accelerators, making particles collide with  unboundedly large energies.
	The objections against this process \cite{Thorn,Jacobson} reduced to the following:  (1) black holes described by the extreme Kerr solution presumably do not exist in nature. At the same time  the energy gain increases    exceedingly  slowly as the maximally spinning case is approached, so the extreme Kerr is a nonadequate description of a realistic BH; (2) high energy collisions take place very close to the horizon and the   time  needed for the outgoing particle to reach a distant observer outside the BH\footnote{This  particle is highly red-shifted due to the energy loss in the  gravitational field of the BH. However, this does not compromise the BH as super collider: such a particle can still be of interest being an outcome  of a superhigh-energy reaction. } is of cosmological scale.

What gave reason to hope that the listed objections could be got around was the fact---discovered by Grib and Pavlov  (GP) in their well-known paper \cite{GP}, see also \cite{GaoZhong},---that there are poles  similar to BSW's in the  \emph{non}-extreme case, too. The problem, however, is that there is a ``potential barrier'' that does not let particles falling  from infinity with suitable parameters (energy and angular momentum) approach the horizon and collide.  The way out proposed by GP (for some generalizations see \cite{Z1,Z2,PatilJoshi} and  references therein) was to employ  additional collisions which would slightly change the parameters of one of the incoming particles, $\alpha$, and enable it to reach the target, the particle $\mu$, in the desired point (see the next section for details). This scenario (complemented with some conjecture on the properties of multiple scattering in accretion disks, see \cite{GP}) has been well received by the community \cite{ZhuWuLiu,revHarKim} and seemed to settle the whole problem.

To summarize, at present the question of whether (nonextreme Kerr) black holes can generate unboundedly high-energy\footnote{Not to be confused with \emph{very} high-energy.}  particle collisions reduces essentially to the question of whether the above mentioned   mechanism is feasible.
In this paper we argue that---in its simplest variant---it is not:
 for particles with bounded energy at infinity  the collision energy remains bounded too (for a rigorous formulation see section~\ref{sec:assert}).

 For the sake of simplicity and definiteness we consider a special case. So, technically speaking one might hope that the unboundedly high acceleration can be achieved by removing the corresponding constraints, i.~e., by
  allowing the particles to leave the equatorial plane or by using a different  way of correcting the trajectory of $\alpha$ (in the general case $\alpha$ is affected by a set of particles. They fall  into the ergosphere, interact  with each other, and some of them hit $\alpha$ (more than once, perhaps).)\footnote{Yet another option is to abandon auxiliary collisions and instead to allow $\alpha$ to decay into two lighter particles \cite{GPm12}. This, though, reduces to our scenario  with $m_2=0$.}. Either of these situations differs from ours, but not \emph{qualitatively}. So, it is hard to imagine why a quantity bounded in the latter case would be \emph{un}bounded in the former. This strongly suggests that the question in the title must be answered negatively in the general case too. Moreover, we conjecture that the same is true for BHs of \emph{any} type as long as they are stationary and lack geodesics infinitely approaching the horizon.

\section{The Kerr black hole as a super accelerator}
Consider
a pair of particles freely falling in a rotating black hole.
  Our subject matter is  the energy $E_{\rm c.m.}$ of these particles'  collision as measured in their centre of mass system.
Can the black hole be a ``super accelerator,'' i.~e. can  $E_{\rm c.m.}$ turn out to be unboundedly large, given initially, in the asymptotically flat region $r\to\infty$, the particles  are (almost) at rest in
an appropriate sense? At first glance the answer is negative:
 it is conceivable that $E_{\rm c.m.}$ unboundedly grows as some parameter $\varsigma$ tends to its limiting value  $\varsigma_0$, but then what happens \emph{at}
      $\varsigma=\varsigma_0$ looks puzzling.
         Surprisingly, the correct answer   \emph{may be positive}. This was discovered by  Banados,  Silk, and West \cite{bsw}
who considered the situation when the black hole in discussion is described by the \emph{extreme} Kerr solution, which is
the case $a=\textcolor[rgb]{0.00,0.00,1.00}{M}$ of the spacetime  \cite[Section 54]{Chandrasekhar}
 \begin{multline}
\label{eq:kerr}
    \rmd s^2 = \rmd t^2 -
\frac{2 M r \, ( \rmd t - a \sin^2 \! \theta\, \rmd \varphi )^2}{r^2 + a^2 \cos^2\theta }
\\
- (r^2 + a^2 \cos^2 \! \theta ) \left( \frac{d r^2}{\Delta} + d \theta^2 \right)
- (r^2 + a^2) \sin^2 \! \theta\, d \varphi^2
\end{multline}
(here
\[
\Delta \equiv r^2 - 2 M r + a^2,
 \]
 while $M$ and $a$ are parameters called the \emph{mass} and \emph{specific angular  momentum} of the space).
 In this spacetime the distance  between the event horizon
\[r = r_H\equiv M + \sqrt{M^2 - a^2}\]
and a point outside it\footnote{Here by ``distance"  the quantity is understood defined, say, as $\int_{r_H}^{r(s)}\sqrt{g_{rr}(r)}\,\rmd r$, where $s$ is the mentioned point.}, is \emph{infinite} in contrast to the \emph{nonextreme} case $a<M$. In particular, there are timelike geodesics outside the horizon  that approach the latter but never reach it (loosely speaking, the corresponding   particle falls towards the horizon for an infinitely long time, no wonder it gains  an infinitely large energy).

 Consider a massive  particle freely falling in the $r > r_H$ part of the  Kerr black hole. Let this particle move in the equatorial plane (from now on the same is required from \emph{all}  particles under consideration). Then it moves  on  a  timelike geodesic $\alpha\equiv\Big(t(\tau),\, r(\tau),\,0,\,  \phi(\tau)\Big)$,
 where  $\tau$ stands for the proper time of the particle, obeying the  conditions \cite[Section 61]{Chandrasekhar}  that
 \begin{equation}\label{eq:e,L}
    \varepsilon \text{ and } L
    \text{ are constant  along }\alpha,
    \end{equation}
 where
\begin{subequations}\label{eq:geod}
\begin{equation}
\varepsilon\obozn g(\partial_t,\partial_\tau )= \dot t \Big(1-\frac{2 M }{r}\Big) +\dot \phi
\frac{2a M }{r},
\end{equation}
\begin{equation}
L\obozn - g(\partial_\phi,\partial_\tau )= -
\frac{2a M }{r}\dot t + \Big[(r^2 +a^2) + \frac{2a^2 M }{r}\Big]    \dot \phi
\end{equation}
 and that
    \begin{equation}\label{def:r}
( r\dot r)^2 =U(\varepsilon, L;r)  \obozn\varepsilon^2 r^2+
\frac{2 M}{r } \, (a  \varepsilon - L)^2 +a^2  \varepsilon^2 - L^2
\  - \Delta,
\end{equation}
\end{subequations}
    where    $\dot{\mathstrut}\obozn \frac{\rmd }{\rmd  \tau}$.
 \begin{teor*}{Remark 1}\label{rem:el}
$\varepsilon $ and $ L$ can be viewed  as characteristics (functions) of a vector ($\partial_\tau  $ in this case), but also, due to \eqref{eq:e,L},   as characteristics of the geodesic, $\alpha$, to which that vector is tangent. In the latter case they are called, respectively, \emph{the specific total energy} (at infinity) and  \emph{the angular momentum} of the geodesic.
The coordinates of a particle at some particular
$\tau_0$  in combination with relations~\eqref{eq:geod} fully determine the trajectory of the particle.
    In particular, the function $U$ may serve as an indicator of accessibility: eq.~\eqref{def:r} implies that a particle with $L=L_0$ and $\varepsilon= \varepsilon_0$ cannot find itself at a point with $r=r_0$ if $U(\varepsilon_0, L_0;r_0)<0$.
    \end{teor*}

Consider a collision  at a point \textcolor[rgb]{0.00,0.00,1.00}{$o$} of two particles of masses $m_\alpha $ and $m_\mu $. Let their  world lines before \textcolor[rgb]{0.00,0.00,1.00}{$o$} be certain  geodesics $\alpha$
  and  $\mu$. In the center of mass system the energy $E_{\rm c.m.}$  of the collision
  is \cite{gen}
      \begin{equation} \label{gen:12}
E_{\rm c.m.}^2(\alpha,\mu; o)= m^2_\alpha+m^2_\mu  +2 m_\alpha m_\mu  g\Bigl(\dot\alpha(o),\dot \mu(o)\Bigr),
\end{equation}
where $\dot\alpha$  and $\dot \mu$ are the velocities at \textcolor[rgb]{0.00,0.00,1.00}{$o$} of $\alpha$
  and  $\mu$, respectively, while $g(\cdot ,\cdot )$ is the scalar product $g(\mathbf{a} ,\mathbf{b})\equiv a_ib^i$.
  $E_{\rm c.m.}$ is a local quantity, so we could alternatively write $E_{\rm c.m.}^2(\dot\mu,\dot\alpha; o)$,  cf.\  remark~1.

  What BSW discovered was a pole in the expression \eqref{gen:12} in the case when particles that initially (at infinity) were at rest, collide at the horizon of an extreme Kerr BH. However,
when a black hole is nonextreme any body reaches the horizon within a finite (proper) time and one does not expect any infinity to appear  as a result of this process. So, it came as a great surprise when
Grib and Pavlov  discovered \cite{GP} that  $E_{\rm c.m.}$ \emph{has poles  even  in the nonextreme case}. \textcolor[rgb]{0.00,0.00,1.00}{
\cite[Eqs.~11,13]{GPm12} say (in our notation)
    \begin{multline}
\frac{E_{\rm c.m.}^{2}(\alpha,\mu; o)}{2\, m_\mu m_\alpha} = \frac{m_\mu^2 + m_\alpha^2}{2 m_\mu m_\alpha} -
\varepsilon_\mu \varepsilon_\alpha +
\frac{1}{r \Delta} \Biggl[ L_\mu L_\alpha (2M-r)
+ 2 \varepsilon_\mu \varepsilon_\alpha \biggl( r^2 (r-M) + a^2 (r+M) -
a M\Bigl( \frac{ L_\mu}{\varepsilon_\mu} + \frac{ L_\alpha}{\varepsilon_\alpha} \Bigr)
\!\biggr)
 \\
- M\sqrt{ 2 \varepsilon_\mu^2 r^2 + 2 (L_\mu - \varepsilon_\mu  a )^2  - L_\mu^2 r/M
+ (\varepsilon_\mu^2  - 1 ) r \Delta/M^2 }
\\ \times
\sqrt{ 2 \varepsilon_\alpha^2 r^2 + 2 (L_\alpha - \varepsilon_\alpha  a )^2  - L_\alpha^2 r/M
+ (\varepsilon_\alpha^2  - 1 ) r \Delta/M^2 }\Biggr]
\label{KerrL1L2}
\end{multline}
whence according  to \cite{GPm12}
\begin{multline}
 E_{\rm c.m.}^{2}\evalat{}{r = r_H}{} =
m_\alpha^2 + m_\mu^2 - \frac{m_\alpha m_\mu}{2M^2}L_{\alpha H} L_{\mu H}
 \\
+ \frac{ m_\alpha m_\mu}{4M^2} \left[ (L_{\alpha H}^2 +4M^2) \frac{L_{\mu H} - L_\mu}{L_{\alpha H} - L_\alpha} +
(L_{\mu H}^2 +4M^2) \frac{L_{\alpha H} - L_\alpha}{L_{\mu H} - L_\mu}
\right],
\label{GPm12:13}
\end{multline}                                      }
    where
\textcolor[rgb]{0.00,0.00,1.00}{\begin{equation}\label{def crit}
    L_{\varpi H}(\varepsilon_\varpi)  \obozn (2   r_HM/a )\varepsilon_\varpi,\qquad \varpi =\alpha,\mu,
\end{equation}
}     and $\varepsilon_\varpi$ is the  specific total energy of the particle $\varpi$. A pair of parameters $\varepsilon$ and $L$ related by the equality \textcolor[rgb]{0.00,0.00,1.00}{$L=(2 r_HM/a)\varepsilon $} is termed \emph{critical}.
It follows from \eqref{GPm12:13} that the energy of a two-particle collision near the horizon  of a Kerr black hole is arbitrarily large if  a particle that has $L=L_1 \neq L_{1H} $ collides with a particle whose parameters
 are  sufficiently close to critical.

\textbf{Warning.}
In considering particle mechanics on the Kerr background one
should be particularly careful with the word ``energy." If we pick a unit timelike vector $\bi T$ at a point $p$ of a geodesic $\alpha(\tau)$, then the quantity $\mathcal E_{\bi T}(\alpha)\obozn m g(\bi T,\dot\alpha)$ is the energy at $p$ of the particle of mass $m$ moving on   $\alpha$ as measured by the observer who is located at $p$ and whose 4-velocity at that point is $\bi T$. This notion is closely related to the ``energy of collision" $E_{\rm c.m.}$ appearing in  \eqref{gen:12}. But there is also
the ``energy at infinity" $m\varepsilon(\alpha)$ of a  particle $\alpha$. The BSW and GP effects show that particles of equal energies $\varepsilon$ may have most different energies $\mathcal E_{\bi T}$.

\textcolor[rgb]{0.00,0.00,1.00}{For
future use note that when
\begin{subequations}\label{eqs:streml}
\begin{equation}\label{eq:L1L2}
\varepsilon_\mu,  L_\mu , \text{are constant}, \quad \varepsilon_\alpha,  L_ \alpha \text{ tend to constants} , \quad L_\mu\neq L_{\mu H}(\varepsilon_\mu)
\end{equation}
and
\begin{equation}\label{eq:->0}
\frac{L_{\alpha}}{ \varepsilon_\alpha}=\frac{ L_{\alpha H}}{ \varepsilon}  + \varphi,\quad\ r(o)=r_H + \xi,
\end{equation}
\end{subequations}
where $L_{\alpha H}$, $\varepsilon$ is a critical pair and 
$\varphi$, and $\xi$ are infinitesimal,
the expression in the square brackets in \eqref{KerrL1L2} divided by $\varepsilon_\mu \varepsilon_\alpha $ has the form
\begin{equation}\label{eq:razl}
A +\xi B +\varphi C-
 \sqrt {D_1 + \xi E_1+ O(\xi^2)}\sqrt {D_2 + \xi E_2 +\varphi F+ G \varphi ^2+
o(\xi)} +
o(\xi),
\end{equation}
where the uppercase letters are some constants.
Comparing this with \eqref{GPm12:13}, which says that for some constant $\Phi$, $ E_{\rm c.m.}^{2}\evalat{}{\xi \to 0}{}\sim \Phi\varphi^{-1}$,  we infer  that
\begin{equation}\label{eq:iz13}
A, D_2,F =0, \qquad D_1 , E_2, G\neq 0, \qquad D_1  G =C^2.
\end{equation}
Consequently, in the leading order the expression \eqref{eq:razl} is equal to
\[\left\{
    \begin{array}{ll}
    \frac{ E_2}{2\sqrt{ G}}\, \xi\varphi^{-1}, & \hbox{at $\xi= o(\varphi^2)$;} \\
     \relax     [C-\sqrt {D_1 (E_2+G)}]\,\sqrt\xi, & \hbox{at $\varphi \sim\sqrt \xi$;} \\
      \sqrt {D_1 E_2}\, \sqrt\xi, & \hbox{at $\varphi= o(\sqrt \xi)$}
\end{array}
  \right.
\]
[that the coefficient in the square brackets (in the second line) is non-zero is seen from \eqref{eq:iz13}]. It follows that
\begin{equation}\label{eq:rasho}
 \text{conditions \eqref{eq:L1L2}, \eqref{eq:->0} imply the divergence\ }   E_{\rm c.m.} (\alpha,\mu; o)\to \infty,
\end{equation}
(which does not follow automatically from \eqref{GPm12:13} the latter being applicable only to the case $\xi= o(\varphi)$).}

Equation \eqref{GPm12:13} does not yet make the black hole a  super accelerator: it may happen that
for some reason particles cannot  both come from the asymptotically flat region. And we shall see that this is the case indeed.
\begin{figure}[t]
\centering
\includegraphics[width=0.37\textwidth]{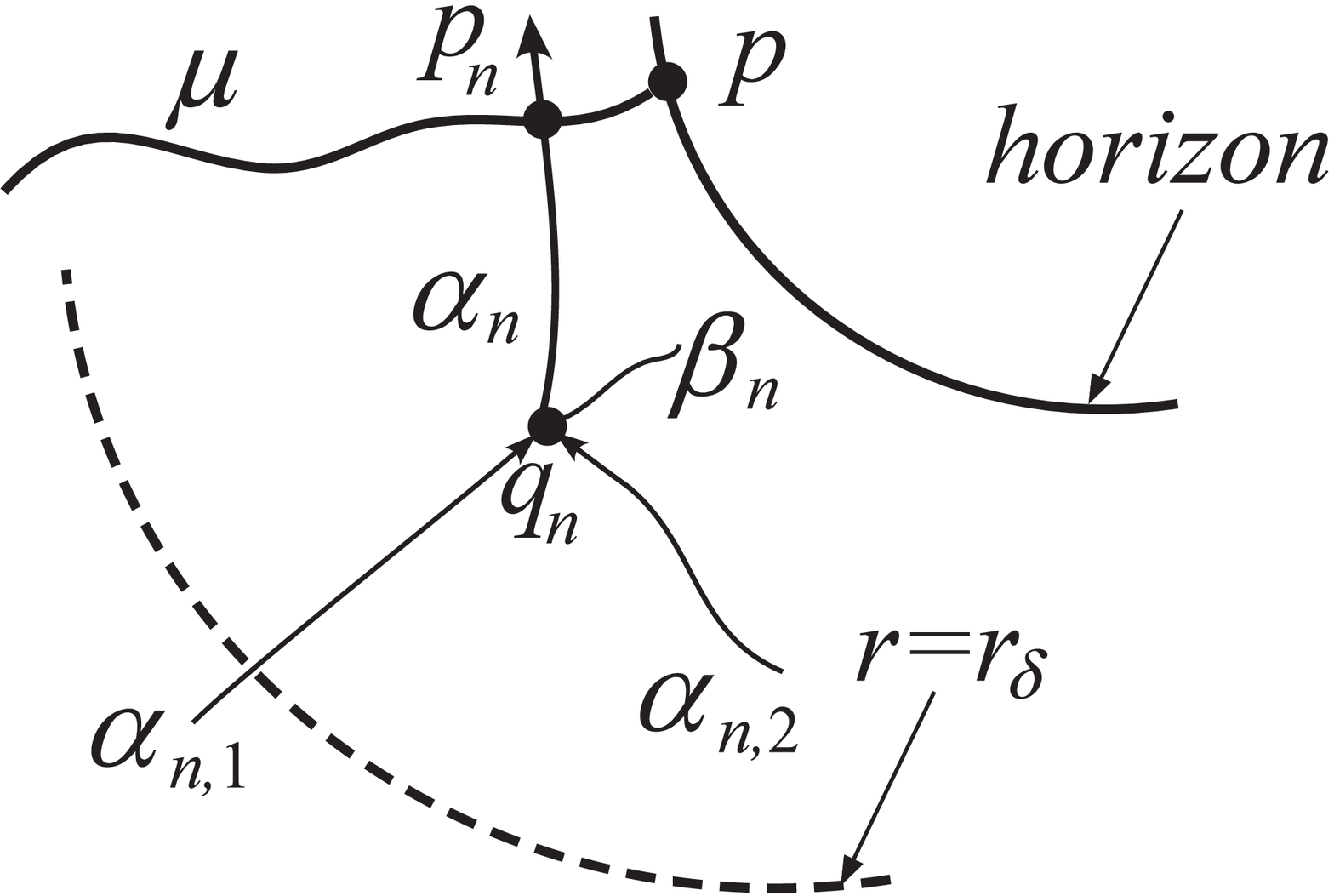}\hfill
\includegraphics[width=0.5\textwidth]{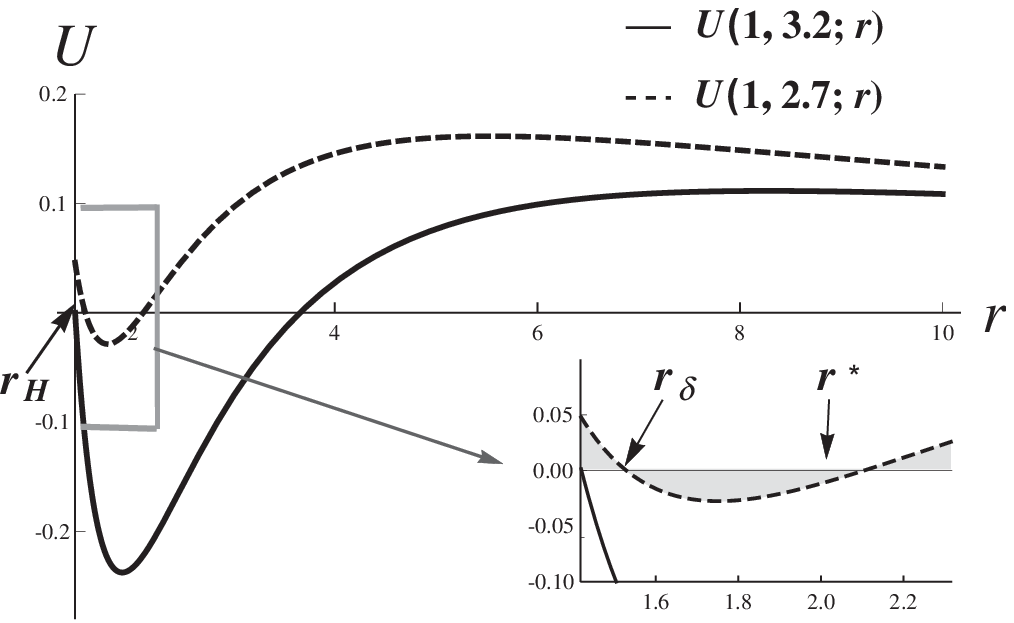}\\
a) \hspace{0.5\textwidth} b)
\caption{a) $\alpha_{n,1}$ and  $\alpha_{n,2}$ come from infinity to collide  \textcolor[rgb]{0.00,0.00,1.00}{at} $q_n:$ $\alpha_{n,1} + \alpha_{n,2}\to \alpha_{n} + \beta_{n}$. \textcolor[rgb]{0.00,0.00,1.00}{One of them collide then with $\mu$ at $p_n$.} b) The function $U(\varepsilon, L;r)$  for  $M=1$, $a=0.9$.
  \label{fig:num} In the case of the solid line the parameters are critical.}
\end{figure}
\textcolor[rgb]{0.00,0.00,1.00}{
Let 
$L_\delta, \varepsilon_\delta$ be functions of $\delta$ such that:
\begin{equation*}
\lim_{\delta\to 0} \varepsilon_\delta = \varepsilon
, \quad \lim_{\delta\to 0} L_\delta= L_H -0  ,
\end{equation*}
where  $L_H$ and $\varepsilon$ is a critical pair. Then\footnote{See, for example, \cite[eq. (25)]{gen}, in which, according to \cite{prCom}, $x_\delta$ is $\tfrac{1}{M}\times$(what we  denote by $r_\delta$).} there exists a
 function $r_\delta > r_H$  such that:
\begin{subequations}\label{eq:rg}
\begin{equation}\label{eq:rg 1}
U (\varepsilon, L;r)\ \text{is non-negative on $[ r_H , r_\delta]$ and changes to negative at } r= r_\delta\\
\end{equation}
 \begin{equation}\label{eq:rg 2}       \lim_{\delta\to 0}r_\delta = r_H
\end{equation}
\end{subequations}
It follows, in particular\footnote{For a rigorous derivation one also must take into consideration that there are no circular geodesics  near the horizon, see the proof of the lemma.}
that  particles  freely falling from infinity \emph{cannot} take the position necessary for colliding with unbounded $E_{\rm c.m.}$: the annulus  $[ r_H , r_\delta(L)]$ that must be entered by the particles is separated from the infinity by the ``potential barrier" $\{U(L)<0\}$.}

To solve this problem in the spirit of GP's ``auxiliary collision" proposal we shall use \emph{pairs} of particles: two particles, $\alpha_{n,1}$  and $\alpha_{n,2}$, see figure~\ref{fig:num}a, are tossed into the ergosphere. They    collide in a point  $q_{n}$ near the horizon and as a result one of them transforms into the particle moving on the geodesic $\alpha_{n}$ with  desired values of the parameters
  $\varepsilon(\alpha_{n}) $ and
 $L(\alpha_{n})\approx L_H(\varepsilon(\alpha_{n}))$.

Though this process leads to unboundedly large  $E_{\rm c.m.}$ it  seems plausible since the difference  $L(\alpha_{n,1}) - L_{H}(\alpha_{n,1})$ before the collision can be made quite small, see \cite{GP}.
This argument, however, is not decisive. It is conceivable, for example,  that the proposed solution suffers exactly the same problem as the initial scenario: to properly correct at  a point $x$ the trajectory of particle 1 the second particle must have parameters $\varepsilon$ and $L$ incompatible with the fact that it came to $x$ from the region outside the  static limit.

In the next section  we show that this is the case and the  above described process  does not allow one to unboundedly accelerate  particles.
 \begin{teor*}{Remark 2}
 Loosely speaking, our result means that the effectiveness $K$ of the nonextreme Kerr black hole as particle accelerator is bounded for any particular energy: \[
 \text{for any }\varepsilon \qquad K(\varepsilon)\equiv\max\frac{E_{\rm c.m.}(\varepsilon)}{\varepsilon}<\infty,\qquad
 \varepsilon\equiv\varepsilon(\alpha_{1}) + \varepsilon(\alpha_{2})
. \]
 \end{teor*}
There is, however, another problem, which is closely related to the former one, but which is too hard to be considered in this paper\footnote{We are grateful to the anonymous referee for drawing our attention to this problem.}. Specifically,  it might  be interesting  to find exactly \emph{how large} $K$ is. In principle, one can imagine that   $\varepsilon=o(E_{\rm c.m.})$, when $\varepsilon\to\infty$, i.~e., that   $K$ of the nonextreme Kerr black holes grows unboundedly with the incoming  energy. This would make such black holes ``almost  super accelerators,'' a phenomenon not yet considered in the literature (to our knowledge).
\section{Unbounded acceleration}\label{sec:assert}

Consider a sequence of elastic collisions realizing our scenario of obtaining a diverging sequence of $E_{\rm c.m.}$. Geometrically speaking, that is a set of points of  the nonextreme Kerr spacetime
 \[q_{n}:\qquad n\in \mathbb N,  \quad\lim_{n\to\infty} r(q_{n})=r_H
 \]
 and timelike future directed geodesic segments $\mu $,  $\alpha_{n,1}$,  $\alpha_{n,2}$, $\alpha_{n}$, $\beta_{n}$---see figure~\ref{fig:num}a---such that:

\textbf{I.}
For each $n$ the geodesics  $\alpha_{n,1}$,   $\alpha_{n,2}$ start at the infinity $r=\infty $ and terminate at the point $q_{n}$, where the geodesics $\alpha_{n}$ and $\beta_{n}$ start. The particles
             $\alpha_{n,1}$ and $\alpha_{n,2}$ have masses $m_1$ and $m_2$, while the masses of the particles moving on
       $\alpha_{n}$ and $ \beta_{n}$ are   $m_3$ and $m_4$, respectively. The geodesics are related by the equality
      \begin{equation}\label{wq:Ep}
        m_1\dot \alpha_{n,1} + m_2\dot\alpha_{n,2}\evalat{}{q_{n}}{}= m_3\dot\alpha_{n} + m_4\dot\beta_{n}
      \end{equation}
    (energy-momentum conservation).

\textbf{II.}
The geodesic $ \mu$ is noncritical, i.~e.,
 \begin{equation}\label{eq: muM}
     L(\mu)=L_\mu\neq L_{\mu H}
   \end{equation}
   and it is the collision of   $ \mu$ with $\alpha_{n}$ that occurs with unboundedly high $E_{\rm c.m.}$:
     \begin{equation}\label{eq:mu divM}
 \textcolor[rgb]{0.00,0.00,1.00}{  \lim_{n\to\infty}g\Bigl(\bi v_n, \dot \alpha_{n}(p_n)\Bigr)={}\infty.
} \end{equation}
   Here    $L_\mu$  is some constant  and  $\bi v_n$  at each $n$ is the  vector of \textcolor[rgb]{0.00,0.00,1.00}{ $  T_{p_n}$}  tangent to the geodesic obtained from $\mu$ by a combination of   isometries $\phi\to \phi+ c_1$ and $t\to t+ c_2$, so all $\bi v_n$ have the same $\varepsilon$ and $L$.

Let us show that in spite of \eqref{eq:mu divM} the particles under discussion are \emph{not} (unboundedly) accelerated by a  nonextreme  black hole: \emph{the diverging $E_{\rm c.m.}$ implies the diverging energies of the incoming  particles $\alpha_{n,i}$.}
\begin{teor*}{Assertion}\label{assert} The  sequence $\{\alpha_{n,i}\}$  contains a   subsequence $\{\alpha_{k,i_0 }\}$, $i_0\equiv 1$ or  $i_0\equiv 2$, with unboundedly large energies at infinity:
\begin{equation}
\lim_{k\to \infty} \varepsilon_{k,i_0}=+\infty
\end{equation}
(from now on we write $L_{n,i}$, $\varepsilon_{n,i}$ for $L(\alpha_{n,i})$, $\varepsilon(\alpha_{n,i})$ and \textcolor[rgb]{0.00,0.00,1.00}{ $L_{n}$, $\varepsilon_{n}$ for $L(\alpha_{n})$, $\varepsilon(\alpha_{n})$}).
 \end{teor*}
 We begin with establishing an  auxiliary---purely algebraic%
 ---property of $U$.
 \begin{teor*}{Lemma}\label{corr}
 For any pair of critical parameters  $\varepsilon^*$  and $L^*$ there is  a constant  $r^*> r_H$ such that
 \begin{equation}\label{corr}
 U(\varepsilon^*, L^*;r)\evalat{}{(r_H,r^*)}{}<0.
\end{equation}
\textcolor[rgb]{0.00,0.00,1.00}{(which means, in particular, that no  critical  geodesic connects the infinity to the horizon)}.  \end{teor*}
 \par\noindent\emph{Proof}\footnote{One can prove the lemma by simply substituting the definition \eqref{def crit} into the definition  of $U$, but the relevant manipulations are rather   formidable.}.
By \eqref{eq:rg}, there  is an $r_\delta$ (converging to $r_H$ when $\delta\to 0$) such that for all sufficiently small positive constants $\delta$ the function $U(\varepsilon^*,L^*-\delta;r)$ is non-negative at $r_H \leq r\leq r_\delta$ and changes its sign at $r_\delta$. Hence,
\begin{equation}\label{eq:U=0}
U(\varepsilon^*,L^*;r_H)=\lim_{\delta\to 0} U(\varepsilon^*,L^*-\delta; r_\delta )=0,
\qquad U'(\varepsilon^*,L^*;r_H)=\lim_{\delta\to 0} U'(\varepsilon^*,L^*-\delta; r_\delta )\leq 0,
\end{equation}
where the prime denotes the derivative by $r$. But \textcolor[rgb]{0.00,0.00,1.00}{if \eqref{corr}  does not hold,} $U$ cannot satisfy the system $U(r_H)= U'(r_H)=0$, because this would mean  that there is a circular orbit \cite{BPT} at $r=r_H$, while such orbits are known to be lacking \cite{Chandrasekhar} in the nonextreme Kerr black hole. So, $U'(\varepsilon^*,L^*;r_H)<0$, cf.~figure~\ref{fig:num}b. Being combined with the first  equality in \eqref{eq:U=0} this proves the lemma.
      \par\nopagebreak \hfill$\square$\par%

 \par\noindent\emph{Proof of the assertion.} To derive a contradiction assume that
 the set \textcolor[rgb]{0.00,0.00,1.00}{$\{ \varepsilon_{n,i}\}$ $i=1,2$, $n=1,2\dots$, \emph{is}} bounded.
Then the set $\{ L_{n,i}\}$ is bounded too,
because at sufficiently large $ L$, $U$ becomes negative [see \eqref{def:r}, at points with $r>2M$]   while such  points  exist  on every $\alpha_{n,i}$ by its very definition. This boundedness implies that for either $i$ there must be  subsequences $\{\varepsilon_{k,i }\}$ and $\{L_{k,i }\}$ converging to some \textcolor[rgb]{0.00,0.00,1.00}{ $\bar\varepsilon_i$  and $\bar L_i$, respectively. Likewise, $\{\varepsilon_{k}\}$, which is the set of the energies of $\{\alpha_{k}\}$, is bounded by $\bar\varepsilon_1 + \bar\varepsilon_2$ and therefore we shall assume without loss  of generality that
$\{\varepsilon_{k}\}$ and $\{L_{k}\}$ converge to some $\bar\varepsilon$  and $\bar L$.
}

\textcolor[rgb]{0.00,0.00,1.00}{By continuity
\[
U(\bar\varepsilon_i,\bar L_i;r)=\lim_{k\to \infty } U(\varepsilon_{k,i }, L_{k,i };r).
\]
But the existence of the geodesics $\alpha_{k,i}$ means that
\[
U(\varepsilon_{k,i }, L_{k,i };r)\evalat{}{[r(q_{k  }),\infty)}{} \geq 0
\]
and we conclude [recall that, by \eqref{eq:rg 2},  $r(q_{k}) \to r_H$] that
\begin{equation*}\label{eq:U>9}
U(\bar\varepsilon_i, \bar L_i;r)\evalat{}{[r_H ,\infty)}{} \geq 0.
\end{equation*}
This  contradicts the lemma and thus completes the proof if
the  parameters  $\bar \varepsilon_i$  and $\bar L_i$ are critical.
}

To prove that this is the case   multiply \eqref{wq:Ep} by $\bi v_k$ to obtain, with the use of \eqref{eq:mu divM}
\begin{equation}\label{eq:1/2}
m_1 g\Big(\bi v_k,\dot\alpha_{k,1}(q_{k})\Big) + m_2 g\Big(\bi v_k,\dot\alpha_{k,2}(q_{k})\Big)  \geq
 m_3g\Big(\bi v_k,\dot\alpha_{k}(q_{k})\Big)
 .
\end{equation}
\textcolor[rgb]{0.00,0.00,1.00}{ By \eqref{gen:12}, the right hand side of \eqref{eq:1/2} equals
\begin{equation}
\frac{1}{2 m_\mu }[E_{\rm c.m.}^2(\alpha_{k},\mu; r(q_{k})) - m^2_3 - m^2_\mu].
\end{equation}
To find the limit of the first term in this expression, note that we come to  the situation of \eqref{eqs:streml} by mere renaming
\[
q_{k}= o,  \quad\varepsilon_k = \varepsilon_\alpha,  \quad
 \bar\varepsilon=\varepsilon, \quad  L_k=L_\alpha,  \quad \bar L=  L_{\alpha H} 
\]
Hence, by \eqref{eq:rasho}, the just mentioned term diverges [it is just to replace $E_{\rm c.m.}(r(q_{k}))\leftrightarrow E_{\rm c.m.}(r(p_{k}))$ that we need the statement \eqref{eqs:streml}].
}
Correspondingly, the left hand side of \eqref{eq:1/2} diverges  too and for $i_0\equiv 1$ or for $i_0\equiv 2$
\begin{equation*}
g\Bigl(\bi v_k,\dot\alpha_{k,i_0 }(q_{k})\Bigr)\xrightarrow[k\to\infty]{}\infty
\end{equation*}
which implies that
\begin{equation*}\label{eq:Ediv}
    E_{\rm c.m.}\Bigl(\bi v_k,\dot\alpha_{k,i_0 }(q_{k})\Bigr)\xrightarrow[k\to\infty]{}\infty.
\end{equation*}
Thus $E_{\rm c.m.}$ has a pole in $\varepsilon=\bar \varepsilon_{i_0}$, $ L=\bar L_{i_0}$   (recall that $\bar \varepsilon_{i}$  and $\bar L_{i}$ were defined as the limits of $\{\varepsilon_{k,i }\}$ and $\{L_{k,i }\}$, respectively). And from \eqref{GPm12:13} it is easy to see that---due to \eqref{eq: muM}---in the poles of  $E_{\rm c.m.}$  the parameters $\varepsilon$ and $L $ are critical indeed.

  \par\nopagebreak \hfill$\square$\par%

\section*{Acknowledgements}
We appreciate kind assistance of Kirill A. Bronnikov. The publication has been prepared with the support of the ``RUDN University Program 5-100''. S. K. is grateful to RFBR for financial support under grant No.~18-02-00461 "Rotating black holes as the sources of particles with high energy."

\end{document}